\documentclass[reprint,amsmath,amssymb,aps,superscriptaddress]{revtex4-2}

\usepackage{graphicx}
\usepackage{subfigure}
\usepackage{comment}
\usepackage{dcolumn}
\usepackage{bm}
\usepackage{amsmath,amssymb}
\usepackage{amsthm}
\usepackage{mathrsfs}
\usepackage{indentfirst}
\usepackage{enumerate}
\usepackage{float}
\usepackage{braket}
\usepackage[utf8]{inputenc}
\usepackage{bm}
\usepackage{graphicx}
\usepackage{tikz}
\usepackage{physics}
\usetikzlibrary{arrows, shapes.gates.logic.US, calc}
\usetikzlibrary{angles,quotes}
\tikzstyle{branch}=[fill, shape=circle, minimum size=3pt, inner sep=0pt]
\usepackage{blochsphere}
\usepackage{mathtools}
\usepackage{color,graphicx}
\usepackage{nccmath}

\usepackage[normalem]{ulem}

\newcommand\D{\!\operatorname{d}\!}
\newcommand{\rh}[1]{\hat{\mathrm{#1}}}
\newcommand{\rr}[1]{\mathrm{#1}}
\newcommand{\cl}[1]{\mathcal{#1}}
\newcommand{\update}[1]{}

\definecolor{C1}{HTML}{ff7f0e}

\newcommand{\mpro}[1]{\textcolor{red}{#1}}

\usepackage{bbold}

\usetikzlibrary{quantikz}

\usepackage{hyperref}
\hypersetup{
 colorlinks=true,
 citecolor=red,
 linkcolor=blue,
 urlcolor=blue,
 pdfpagemode=UseNone,
 pdfstartview=FitH}

\usepackage{subfiles}

\newcommand{\orcidicon}[1]{\href{https://orcid.org/#1}{\includegraphics[height=\fontcharht\font`\B]{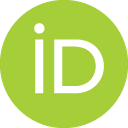}}}

\footnotetext{These authors contributed equally}

\begin{document}
\preprint{APS/123-QED}
\title{Simulating photonic devices with noisy optical elements}
\author{Michele Vischi\,\orcidicon{0000-0002-5724-7421}$^\diamond$}

\email{michele.vischi@phd.units.it}
\affiliation{Department of Physics, University of Trieste, Strada Costiera 11, 34151 Trieste, Italy}
\affiliation{Istituto Nazionale di Fisica Nucleare, Trieste Section, Via Valerio 2, 34127 Trieste, Italy}
\author{Giovanni Di Bartolomeo\,\orcidicon{0000-0002-1792-7043}$^\diamond$}
\email{giovanni.dibartolomeo@phd.units.it}
\affiliation{Department of Physics, University of Trieste, Strada Costiera 11, 34151 Trieste, Italy}
\affiliation{Istituto Nazionale di Fisica Nucleare, Trieste Section, Via Valerio 2, 34127 Trieste, Italy}
\author{Massimiliano Proietti}
\affiliation{Leonardo Labs, Quantum Technologies Lab, Via Tiburtina, KM 12,400 - Rome - 00131 - Italy
}
\author{Seid Koudia}
\affiliation{Leonardo Labs, Quantum Technologies Lab, Via Tiburtina, KM 12,400 - Rome - 00131 - Italy
}
\author{Filippo Cerocchi}
\affiliation{
Leonardo, Cyber \& Security Solutions Division, Via Laurentina, 760 - 00143 Rome - Italy
}
\author{Massimiliano Dispenza}
\affiliation{Leonardo Labs, Quantum Technologies Lab, Via Tiburtina, KM 12,400 - Rome - 00131 - Italy
}
\author{Angelo Bassi\,\orcidicon{0000-0001-7500-387X}}
\affiliation{Department of Physics, University of Trieste, Strada Costiera 11, 34151 Trieste, Italy}
\affiliation{Istituto Nazionale di Fisica Nucleare, Trieste Section, Via Valerio 2, 34127 Trieste, Italy}
\begin{abstract}
Quantum computers are inherently affected by noise. While in the long-term error correction codes will account for noise at the cost of increasing physical qubits, in the near-term the performance of any quantum algorithm should be tested and simulated in the presence of noise. As noise acts on the hardware, the classical simulation of a quantum algorithm should not be agnostic on the platform used for the computation. In this work, we apply the recently proposed noisy gates approach to efficiently simulate noisy optical circuits described in the dual rail framework. The evolution of the state vector is simulated directly, without requiring the mapping to the density matrix framework. Notably, we test the method on both the gate-based and measurement-based quantum computing models, showing that the approach is very versatile. We also evaluate the performance of a photonic variational quantum algorithm to solve the MAX-$2$-CUT problem. In particular we design and simulate an ansatz which is resilient to photon losses up to $\rr{p} \sim 10^{-3}$ making it relevant for near term applications. 
\end{abstract}
\maketitle
\section{Introduction}\label{intro}
Photonic quantum devices are very promising potential candidates for achieving quantum advantage \cite{madsen2022quantum,zhong2020quantum}: integrated photonic benefits from low error rates and ease of manipulation, making it eligible for scalable quantum computation \cite{mezher2022assessing, maring2023general}. However, like all quantum computing platforms, photonic devices in the NISQ era are affected by noise from various sources, such as photon losses and imperfect indistinguishability of single photons, that can negatively impact the performance of quantum algorithms \cite{mezher2022assessing, pont2022quantifying, czerwinski2022statistical}. Therefore, studying the noise effects on quantum algorithms and perform accurate classical simulations becomes crucial for their successful deployment \cite{niedermeier2023tensor, wright2022numerical, cheng2021simulating}. This task can be achieved by employing efficient classical simulation methods based on reliable  noise model. 


Building on the results presented in~\cite{dibartolomeo2023novel}, we develop the noisy gates approach for optical devices in the dual-rail encoding. The distinctive feature of this approach is that the noise is integrated in the dynamics defining the quantum gates---which turn into noisy gates---offering in this way a superior tool for simulating the behavior of a (unavoidably) noisy quantum device. Here we extend the formalism to second quantization for optical elements \cite{kok2006linear,knill2000efficient}, and apply it to specific instances of dual-rail encoding, showing that it is a very flexible and efficient framework for simulating noisy optical computers. 

We successfully formulate the noisy versions of linear optical elements as phase shifters and beam splitters; we also consider the effects of imperfect single-photon sources, lossy optical guides and detectors as fictitious noisy optical elements that act on photons at the proper step of the circuit. The approach is first tested through simulations of both gate based and measurement based quantum circuits, whereby the same algorithm is expressed in the dual-rail formalism. Then, the approach is applied to a specific variational algorithm solving an optimisation problem, through the study of noise effects on the optimization loop of a variational quantum algorithm. 

The paper is organized as follows. In Sec. \ref{formalism} we recap the nosy gates formalism. In Sec. \ref{noise_dual_rail} we review the main sources of noise in optical devices. Sec. \ref{noisy_optical_elements} and \ref{other_noises} present the general derivation of noisy optical elements, imperfect single-photon source effects, lossy optical guides and detectors. In Sec. \ref{comparisons} we compare our method with previuos existing ones. In Sec. \ref{tests} we report the results of the simulations of X gate and Bell state preparation circuit in the gate based quantum computing framework, of the X gate in the measurement based framework and of a variational quantum algorithm applied to the max 2-cut problem. We conclude with some general remarks and an outlook.

\section{The Noisy gates formalism}\label{formalism}
Noise models for quantum computing typically consists in adding extra gates, mimicking the noise, before and after the standard gates defining a circuit \cite{benenti2019principles, georgopoulos2021modeling,rigetti_noise}.
The noise gate approach developed in~\cite{dibartolomeo2023novel} instead combines the gate and the noise into a single noisy gate, which represents a more realistic description of the actual functioning of a noisy quantum device. As a matter of fact, this approach proved to offer a better simulation of superconducting quantum computers. We briefly present its building blocks. 


Noises affect quantum devices and their effects can be described by the theory of open quantum systems \cite{breuer2002theory,nielsen2000quantum}. The master equation that describes the Markovian evolution of the density matrix of the qubits system in a given environment is
\begin{equation}
\label{mastereq}
\frac{\D}{\D t} \rh{\rho}_t = -\frac{i}{\hbar}[\rh{H}_t,\hat{\rho}_t] + \epsilon^{2} \sum_{k}\biggl[\rh{L}_{k}\rh{\rho}_t\rh{L^{\dagger}}_{k} - \frac{1}{2}\{\rh{L^{\dagger}}_{k}\rh{L}_{k},\rh{\rho}_t\}\biggr],
\end{equation}
where $\rh{H}_t$ is the time dependent Hamiltonian realizing a given gate, and $\rh{L}_{k}$ are the Lindblad operators capturing the action of the environment. In principle, in Eq.\eqref{mastereq}  each term $\rh{L}_{k}$ should correspond to a different parameter $\epsilon^2_k$; however we assume that the noise strengths are of the same order of magnitude, thus for our purposes we consider a single parameter $\epsilon^2$. 

Given that $\epsilon^{2}$ is much weaker than the  strength of the Hamiltonian, one usually assumes that the noises are totally independent from the gates dynamics, allowing to formally decouple the two terms in Eq. (\ref{mastereq}). In \cite{dibartolomeo2023novel} we showed that the assumption underlying this approximation is too tight and one can obtain better results by solving a stochastic differential equation, equivalent to Eq.~\eqref{mastereq}, perturbatively to the second order. The approach is the following: we perform a linear stochastic unraveling of the Lindbald equation~\cite{jacobs1998linear,caiaffa2017stochastic,bassi2003stochastic,jacobs2014quantum,wiseman2009quantum}, in terms of the following stochastic differential equation for the statevector
\begin{equation}
\label{SDE}
\D \ket{\psi_t} = \biggl[-\frac{i}{\hbar}\rh{H}_t\D t +  \sum_{k} \biggl(i\epsilon\rh{L}_{k}\D \rr{W}_{k}(t)- \frac{\epsilon^{2}}{2}\rh{L^{\dagger}}_{k}\rh{L}_{k} \D t\biggr)\biggr]\ket{\psi_t} \, ,
\end{equation}
where $\rr{d}\rr{W}_{k}(t)$ are differentials of standard independent Wiener processes, i.e. stochastic infinitesimal increments such that $\mathbb{E}\big[\rr{d}\rr{W}_{k}(t)\big]=0$ and $\mathbb{E}\big[ \rr{d}\rr{W}_{k}(t)\rr{d}\rr{W}_{k}(t')\big]=\delta_{k,k'}\rr{d}t $. Eq.~\eqref{SDE} is an unraveling of the Lindblad equation in the sense that the density matrix obtained by averaging the pure states $\ket{\psi_t}\bra{\psi_t}$ over the noise
\begin{equation}\label{averaged_solution}
    \rh{\rho}_t=\mathbb{E}\Big[  \ket{\psi_t}\bra{\psi_t}  \Big], 
\end{equation}
is a solution of Eq.~\eqref{mastereq}. In this sense, Eqs.~\eqref{mastereq} and ~\eqref{SDE} have the same physical content.

Moreover Eq.~\eqref{SDE} is linear, thus the solution can be written as
\begin{equation}
\ket{\psi_t} = \rh{N}_{g}(t,t_{0})\ket{\psi_{t_{0}}},
\end{equation}
where $\rh{N}_{g}(t,t_{0})$ is a linear operator, implying that it can be interpreted as a noisy quantum gate 
\begin{equation*}
\begin{quantikz}
\lstick{$\ket{\psi}$} &\qw &\gate{\rr{N}_{g}} &\qw&\rstick{$\ket{\psi'}$} \qw
\end{quantikz}.
\end{equation*}

In \cite{dibartolomeo2023novel} we showed how to solve Eq. \eqref{SDE} approximately to order $O(\epsilon^2)$ by using perturbative methods for stochastic differential equations \cite{gardiner1985handbook}. The resulting expression for the noisy gates is the following
\begin{equation}\label{noisy_gate}
    \rh{N}_g(t,t_0)=\rh{U}_g(t,t_0)e^{-\frac{\epsilon^{2}}{2}\rh{D}(t,t_0)}e^{i\epsilon\rh{S}(t,t_0)},
\end{equation}
where we defined the deterministic operator
\begin{align}
&\rh{D}(t,t_0) = \sum_{k}\rh{D}_k(t,t_0)\label{D},\\
& \rh{D}_k(t,t_0) = \int_{t_{0}}^{t}\rr{d}s\bigg( \rh{L}^\dag_{k}(s)\rh{L}_{k}(s) -\rh{L}^2_{k}(s) \bigg) \label{Dk}
\end{align}
and the stochastic one
\begin{align}
&\rh{S}(t,t_0) =\sum_{k}\rh{S}_{k}(t,t_0)\label{S},\\
&\rh{S}_{k}(t,t_0) = \int_{t_{0}}^{t}\rr{d\rr{W}_{k}(s)}\rh{L}_{k}(s).\label{Sk}
\end{align}
We notice that in Eq.~\eqref{noisy_gate} the Lindblad operators $\rh{L}_{k}(s)$ are in the interaction picture, $\rh{L}_{k}(s) = \rh{U}^{\dagger}_g(s,t_{0})\rh{L}_{k}\rh{U}_g(s,t_{0})$, thus since they depend on the gate operator, the noise dynamics is influenced by the unitary evolution realized by $\rh{U}_g$. Moreover, the real part $[\rh{S}^{\text \tiny{R}}_{k}(t,t_{0})]_{ij}$ and the imaginary part $[\rh{S}^{\text \tiny{I}}_{k}(t,t_{0})]_{ij}$ of the entries of the operators $\rh{S}_{k}$ in Eq.~\eqref{Sk} are It\^o integrals of deterministic functions: $[\rh{S}^{m}_{k}(t,t_{0})]_{ij} = \int_{t_{0}}^{t} \D\rr{W}_{k}(s)[\rh{L}^{m}_{k}(s)]_{ij}$ for $m = \text \tiny{R} ,\tiny{I}$.  The latter represent Gaussian stochastic processes with means zero $\mathbb{E}\big[[\rh{S}^{m}_{k}(t,t_{0})]_{ij}\big] = 0$, variances $\mathbb{V}\bigl[[\rh{S}^{m}_{k}(t,t_{0})]_{ij}\bigr] = \int_{t_{0}}^{t} \D s([\rh{L}^{m}_{k}(s)]_{ij})^2$ and covariances $\mathbb{E}\bigl[[\rh{S}^{m}_{k}(t,t_{0})]_{ij} [\rh{S}^{n}_{k}(t,t_{0})]_{i'j'}\bigr] =  \int_{t_{0}}^{t} \D s[\rh{L}^{m}_{k}(s)]_{ij}[\rh{L}^{n}_{k}(s)]_{i'j'}$.
Once all the variances and covariances are computed, the stochastic processes can be sampled to generate a single realization of the noisy gate $\rh{N}_g$. By averaging over the number $\rr{N}_{\text{\tiny samples}}$ of realizations of the final state $\ket{\psi_t}_k = \rh{N}_g \ket{\psi_{t_0}}$, one recovers the final density matrix of the system: 
\begin{equation}
\label{mean_rho}
    \rh{\rho} = \frac{1}{\rr{N}_{\text{\tiny samples}}} \sum_k \ket{\psi_t}_k\bra{\psi_t}_k.
\end{equation}
It is then sufficient to specify the Hamiltonians, i.e. the unitaries, and the Lindblad operators of a given quantum device to find the expressions of the noisy gates. In the following we specialize to linear optical elements.

\section{Noises in dual-rail encoding optical circuits}\label{noise_dual_rail}
This work focuses on discrete variable quantum optics \cite{scully1999quantum,vogel2006quantum}, in particular on the so called dual-rail encoding of linear optics quantum computing \cite{burgarth2013dual,burgarth2005conclusive}, where the computational units are qubits. We consider optical circuits that are essentially made up of single photon sources, single-mode optical guides (also called spatial modes) in which photons propagate, linear optical elements that modify the photons state and single photon detectors. Thus, an optical circuit can be thought as a unitary transformation on $N$ modes with symmetry group $SU(N)$, followed by the detection. Since a qubit has a $SU(2)$ symmetry, then two spatial optical modes (dual-rail) with a single photon is a natural implementation of a qubit. 

The logical states of the qubit are encoded in the two modes Fock states $\ket{0}_{L} = \ket{10}$, $\ket{1}_{L} = \ket{01}$ (with this notation we indicate that one of the two modes is occupied by the photon). Thus to represent a system of $M$ qubits one needs $2M$ modes plus a certain number of additional ancillary modes used to implement the correct unitary transformation.

Although it is believed that optical quantum computing is very promising because of the weak interaction of photons with their surrounding environment, optical quantum circuits are not immune to noise. 
We focus on two main sources of noise affecting optical devices: photon losses and non-perfect indistinguishability of single photons \cite{mezher2022assessing,pont2022quantifying}.

\textit{Photon losses}  are particularly daunting in the dual rail encoding, since after a loss the photon state exits from the logical computational states of the single qubit. They can occur at any step of the optical circuit, from the source to optical guides, optical elements and detectors.  For each of these steps, one can associate a different loss probability $\rr{p} = 1 - e^{-\Delta t/T}$, where $\Delta t$ is the time interval and $T$ is a characteristic time. Indeed the state of a photonic mode traveling through a fiber can be expressed as \cite{czerwinski2022statistical,czerwinski2022efficiency,mielczarek2018quantum}
    \begin{equation}
        \hat{\rho}(\Delta t) = \rr{p} \ket{0}\bra{0} + (1-\rr{p})\ket{1}\bra{1}
    \end{equation}
    where $\ket{0}$ is the state of a single mode without photons and $\ket{1}$ is the state of a single mode occupied by the photon.  In practice, the value of p depends specifically on the material used for the optical circuit as well as other experimental parameters. In the following, for the purposes of this work, we vary $\rr{p}$ from $10^{-4}$ to $10^{-2}$.
    
\textit{Photons non-perfect indistinguishability} occurs at the stage of photon/qubits generation. Single photons emitted for example by quantum dots \cite{jacak2013quantum} are indistinguishable if they have the same wavelength, polarization, temporal and spatial extent. A quantum dot sources indistinguishable photons if it is subject only to radiative decay. If there is any other decoherence process, then sourced photons will be not perfectly indistinguishable. For this reason we simulate indistinguishably using decoherence models.
    Typical decoherence processes in semiconductor platforms are depolarization or dephasing \cite{cogan2018depolarization,schneiderdephasing}
    and for this reason somehow photons non-perfect indistinguishability is connected to depolarization/dephasing errors because photons are in turn coupled to an electronic spin state that is susceptible to such errors.

We notice that photon losses and  non-perfect indistinguishability lead to incoherent evolution. Here we do not consider coherent errors caused by imperfect calibration of optical elements. The latter can be accounted for using already existing techniques \cite{pai2019matrix}. Other sources of error, e.g. cross-talk effects, are beyond the scope of this work and can be added with further developments.

\section{Noisy optical elements}\label{noisy_optical_elements}
We now apply the theory outlined in the previous sections to build the noisy version of linear optical elements. In particular we focus on phase shifters and beam splitters, showing how to write down their noisy version under the effect of photon loss errors. 

\subsection{Noiseless phase shifter and beam splitter}
We briefly recap the noiseless expression of optical elements in order to set the main notation used in the following sections.
A phase shifter is an optical element that acts on a single photon mode by changing the phase of the photon state. Its action can be described by the Hamiltonian \cite{kok2006linear,knill2000efficient}
\begin{equation}
\label{phase_H}
\rh{H}_{\rr{P}} = -\hbar \omega  \rh{P},\qquad \rh{P} = \hat{a}^{\dagger}\hat{a},
\end{equation}
where $\hat{a}$ and $\hat{a^{\dagger}}$ are respectively the annihilation and creation operators of a photon in a single mode. By solving the Schr\"odinger equation one arrives at the unitary evolution operator $\rh{U}_{\rr{P}} = e^{i\theta \rh{P}}$,
where $\theta = \omega\cdot (t-t_0)= \omega\cdot \Delta t$. Thus, the evolution of the annihilation and creation operators is given by 
\begin{equation}
\label{phase_relations}
\rh{U}_{\rr{P}}\hat{a}^{\dagger}\rh{U}_{\rr{P}}^{\dagger} = e^{i\theta}\hat{a}^{\dagger}, \qquad
\rh{U}_{\rr{P}}\hat{a}\rh{U}_{\rr{P}}^{\dagger} = e^{-i\theta}\hat{a}.
\end{equation}

The beam splitter acts on two photon modes
and can be defined through the Hamiltonian \cite{kok2006linear,knill2000efficient}
\begin{equation}
\label{beam_H}
\rh{H}_{\rr{B}} = - \frac{\hbar\omega}{2}  \rh{B}(\phi),\quad \rh{B}(\phi) = e^{-i\phi}\hat{a}_{0}\hat{a}_{1}^{\dagger} + e^{i\phi}\hat{a}_{0}^{\dagger}\hat{a}_{1},
\end{equation}
where the indices of the annihilation and creation operators refer to photon modes 0 and 1. 
The unitary evolution operator is $\rh{U}_{\rr{B}} = e^{i\frac{\theta}{2} \rh{B}(\phi)}$,
with $\theta = \omega \Delta t$; the annihilation/creation operators evolve as follows:
\begin{align}
&\rh{U}_{\rr{B}}\hat{a}_{0}^{\dagger}\rh{U}_{\rr{B}}^{\dagger} = \cos{(\theta/2)}\hat{a}_{0}^{\dagger} + ie^{-i\phi} \sin{(\theta/2)}\hat{a}_{1}^{\dagger}, \label{beam_relation_1}\\
&\rh{U}_{\rr{B}}\hat{a}_{1}^{\dagger}\rh{U}_{\rr{B}}^{\dagger} = ie^{i\phi} \sin{(\theta/2)}\hat{a}_{0}^{\dagger} + \cos{(\theta/2)}\hat{a}_{1}^{\dagger},\label{beam_relation_2} \\
&\rh{U}_{\rr{B}}\hat{a}_{0}\rh{U}_{\rr{B}}^{\dagger} = \cos{(\theta/2)}\hat{a}_{0} - ie^{i\phi} \sin{(\theta/2)}\hat{a}_{1},\label{beam_relation_3} \\
&\rh{U}_{\rr{B}}\hat{a}_{1}\rh{U}_{\rr{B}}^{\dagger} = -ie^{-i\phi} \sin{(\theta/2)}\hat{a}_{0} + \cos{(\theta/2)}\hat{a}_{1}. \label{beam_relation_4}
\end{align}
It is possible to express in a compact way these latter transformations by defining unitary matrices that act on vectors of creation operators $(\hat{a}_{0}^{\dagger},\hat{a}_{1}^{\dagger})^{T}$ or annihilation operators $(\hat{a}_{0},\hat{a}_{1})^{T}$. The matrix for creation operators reads
\begin{equation}
\label{U_beam}
\rh{\mathcal{U}}_{\rr{B}} = \begin{pmatrix} \cos{(\theta/2)} & ie^{-i\phi} \sin{(\theta/2)}\\
ie^{i\phi} \sin{(\theta/2)}&\cos{(\theta/2)}
\end{pmatrix},
\end{equation}
while that for annihilation operators is $\rh{\mathcal{U}}_{\rr{B}}^{\dagger}$. 

\subsection{Noisy phase shifter}
A natural way to describe the effects of photon losses on a given mode $0$ in the noisy gates formalism is to consider the system in contact with an environment that absorbs photons. By following the protocol described in Sec. \ref{formalism} we have to choose the operators in Eq. \eqref{SDE}: the Hamiltonian operator is the phase shifter Hamiltonian in Eq. \eqref{phase_H} on mode 0, i.e. $\rh{P}_{0} = \hat{a}_{0}^{\dagger}\hat{a}_{0}$, while  we choose $\hat{a}_{0}$ as Lindblad operator $\rh{L}$.
The corresponding expression for the noisy gate reads
\begin{equation}\label{noisy_phase_shifter_fail}
    \rh{N}_{0}=\rh{U}_{0}e^{-\frac{\epsilon^2_{0}}{2}\rh{D}_{0}}e^{i\epsilon_{0}\rh{S}_{0}} ,
\end{equation}
where $\rh{U}_{0} = e^{i\theta \rh{P}_{0}}$, $\rh{D}_{0} = \int_{t_{0}}^{t}\rr{ds}(\hat{a}^{\dagger}_{0}(s)\hat{a}_{0}(s)-\hat{a}^2_{0}(s))$ and $\rh{S}_{0} = \int_{t_{0}}^{t}\rr{d\rr{W}(s)}\hat{a}_{0}(s)$. 
The time dependent creation operator reads
\begin{equation}
  \hat{a}^{\dagger}_{0}(s) =  \rh{U}^{\dagger}_{0}(s,t_0)\hat{a}^{\dagger}_{0}\rh{U}_{0}(s,t_0) = e^{-i\theta(s,t_0)} \hat{a}^{\dagger}_{0}
\end{equation}
similarly to Eq.\eqref{phase_relations}. Thus we can write
\begin{equation}
\begin{aligned}
& \rh{D}_{0} = \Delta t \hat{a}^{\dagger}_{0}\hat{a}_{0}  +(\rr{D}_{C} -i \rr{D}_{S} )\hat{a}^2_{0}\, ,
& \rh{S}_{0} = (\rr{I}_{C} -i \rr{I}_{S} )\hat{a}_{0}
\end{aligned}
\end{equation} 
where $\rr{D}_{C} = \int_{t_{0}}^{t}\rr{ds}\cos{2\theta(s,t_{0})}$, $\rr{D}_{S} = \int_{t_{0}}^{t}\rr{ds}\sin{2\theta(s,t_{0})}$, $\rr{I}_{C} = \int_{t_{0}}^{t}\rr{d\rr{W}(s)}\cos{\theta(s,t_{0})}$, $\rr{I}_{S} = \int_{t_{0}}^{t}\rr{d\rr{W}(s)}\sin{\theta(s,t_{0})}$. 
The latters are Gaussian stochastic processes with means $\mathbb{E}\big[\rr{I}_{C}\big] = \mathbb{E}\big[\rr{I}_{S}\big] = 0$, variances $\mathbb{V}\bigl[\rr{I}_{C}\bigr] = \int_{t_{0}}^{t} \D s \cos^{2}{\theta(s,t_{0})}$, $\mathbb{V}\bigl[\rr{I}_{S}\bigr] = \int_{t_{0}}^{t} \D s \sin^{2}{\theta(s,t_{0})}$ and covariance $\mathbb{E}\big[\rr{I}_{C}\rr{I}_{S}\big] = \int_{t_{0}}^{t} \D s \cos{\theta(s,t_{0})}\sin{\theta(s,t_{0})}$.

However, at this point a difficulty arises because $\rh{N}_{0} \hat{a}^{\dagger}_{0} \rh{N}^{\dagger}_{0} $ cannot be expressed explicitly as a finite function of the creation operator only.

In order to avoid such problem we consider a larger physical system consisting of two photon modes: mode 0 is the physical mode where the phase shifter acts, while mode 1 is a virtual mode that we exploit to model photon losses. The idea is to model the environment  as a beam splitter between the two modes, and then trace out the virtual mode. In such a way, the effective action on the system is that the photon is absorbed with some probability. In this new configuration,   we choose $\rh{B}_{01} = \hat{a}_{0}\hat{a}_{1}^{\dagger} + \hat{a}_{0}^{\dagger}\hat{a}_{1}$ as Lindblad operator $\rh{L}$; the latter can be obtained from the expression for $\rh{B}(\phi)$ in Eq. \eqref{beam_H} by setting $\phi = 0$. This Lindblad operator acts as a virtual beam splitter that can shift photons from mode 0 to the virtual mode 1.

Following Sec. \ref{formalism}, the resulting noisy phase shifter is of the form in Eq. \eqref{noisy_gate} without the deterministic term due to the fact that the Lindblad operator $\rh{B}_{01}$ that we chose is Hermitian (see also Eq.\eqref{Dk}). Then we can write
\begin{equation}\label{noisy_phase_shifter}
    \rh{N}_{01}=\rh{U}_{0}e^{i\epsilon_{0}\rh{S}_{01}},
\end{equation}
 where $\rh{U}_{0} = e^{i\theta \rh{P}_{0}}$ and $\rh{S}_{01} = \int_{t_{0}}^{t}\rr{d\rr{W}(s)}\rh{B}_{01}(s)$. 
 
 At this point we need to compute the form of the stochastic operator $\rh{S}_{01}$. We notice that
\begin{equation}\label{B_phase_shifter_time}
\begin{aligned}
&\rh{B}_{01}(s) = \rh{U}_{0}^{\dagger}(s,t_{0})\rh{B}_{01}\rh{U}_{0}(s,t_{0})\\
&= e^{-i\theta(s,t_{0})\rh{P}_{0}}\hat{a}_{0}e^{i\theta(s,t_{0})\rh{P}_{0}}\hat{a}_{1}^{\dagger} + e^{-i\theta(s,t_{0})\rh{P}_{0}}\hat{a}_{0}^{\dagger}e^{i\theta(s,t_{0})\rh{P}_{0}}\hat{a}_{1} \\
& = e^{i\theta(s,t_{0})}\hat{a}_{0}\hat{a}_{1}^{\dagger} + e^{-i\theta(s,t_{0})}\hat{a}_{0}^{\dagger}\hat{a}_{1} \\
&= \cos{\theta(s,t_{0})}(\hat{a}_{0}\hat{a}_{1}^{\dagger} + \hat{a}_{0}^{\dagger}\hat{a}_{1}) + i \sin{\theta(s,t_{0})}(\hat{a}_{0}\hat{a}_{1}^{\dagger} - \hat{a}_{0}^{\dagger}\hat{a}_{1}) \\
& = \cos{\theta(s,t_{0})}\rh{B}_{01} - \sin{\theta(s,t_{0})}\rh{C}_{01}
\end{aligned}
\end{equation}
where in the third line we used Eq.\eqref{phase_relations} with the adjoint unitary evolution and we define $\rh{C}_{01} = i(\hat{a}_{0}^{\dagger}\hat{a}_{1} - \hat{a}_{0}\hat{a}_{1}^{\dagger})$, which can be obtained from $\rh{B}(\phi)$ in Eq.\eqref{beam_H} by choosing $\phi = \pi/2$.

Then the stochastic operator reads
\begin{equation}
\rh{S}_{01} = \rr{I}_{C}\rh{B}_{01} - \rr{I}_{S}\rh{C}_{01},
\end{equation}
where $\rr{I}_{C} = \int_{t_{0}}^{t}\rr{d\rr{W}(s)}\cos{\theta(s,t_{0})}$ and $\rr{I}_{S} = \int_{t_{0}}^{t}\rr{d\rr{W}(s)}\sin{\theta(s,t_{0})}$. The latters are the same stochastic integrals defined before.

Finally, by exploiting the fact that at second order in $\epsilon$ one has $e^{\epsilon(\rh{A} + \rh{B})} = e^{\epsilon\rh{A}} e^{\epsilon\rh{B}}$ , from Eq. \eqref{noisy_phase_shifter} we get
\begin{equation}\label{noisy_phase_shifter_2}
    \rh{N}_{01}=\rh{U}_{0}e^{i\epsilon_{0}\rr{I}_{C}\rh{B}_{01}}e^{-i\epsilon_{0}\rr{I}_{S}\rh{C}_{01}}.
\end{equation}
Through Eq. \eqref{noisy_phase_shifter_2} one can compute the evolution of creation and annihilation operators under the action of a noisy phase shifter.  As seen before, we can express compactly these transformations with unitary matrices that act on vectors of creation operators. In the case of the noisy phase shifter we have
\begin{equation}
\label{N_phase}
\begin{aligned}
&\rh{\mathcal{N}}_{01} = \begin{pmatrix} e^{i\theta} & 0\\
0&1
\end{pmatrix}\cdot
\begin{pmatrix} \cos{(\epsilon_{0}\rr{I_{C}})} & i\sin{(\epsilon_{0}\rr{I_{C}})}\\
i\sin{(\epsilon_{0}\rr{I_{C}})}&\cos{(\epsilon_{0}\rr{I_{C}})}
\end{pmatrix}\cdot \\
&\begin{pmatrix} \cos{(\epsilon_{0}\rr{I_{S}})} & -\sin{(\epsilon_{0}\rr{I_{S}})}\\
\sin{(\epsilon_{0}\rr{I_{S}})}&\cos{(\epsilon_{0}\rr{I_{S}})}
\end{pmatrix} \, ,
\end{aligned}
\end{equation}
whereas $\rh{\mathcal{N}}^{\dagger}_{01}$ acts on annihilation operators. In Eq. \eqref{N_phase} the parameter $\epsilon_{0}$ is connected to photon loss probability $\rr{p_{0}}$ on mode 0 as $\epsilon_{0} = \sqrt{-\log{(1-2\rr{p}_{0})}/2}$ where $\rr{p}_{0} = (1 - e^{-2\Delta t/T})/2$ and $\epsilon_{0}^{2} = \Delta t/T$, with $T$ the characteristic time of photon losses. We choose such time dependence for $\rr{p}_{0}$ because by taking the average over the processes $\rr{I}_C$ and $\rr{I}_S$ one has
$\mathbb{E}[\sin^{2}(\epsilon\rr{I}_k)] = (1-e^{-2\epsilon_0^{2}})/2 = \rr{p}_0$ with $k = C,S$. Moreover, we are not interested in the regime in which there is more than 50\%\ of loss probability. 

By tracing out the virtual mode we get the expression for the noisy phase shifter
\begin{equation}
\label{N_phase_trace}
\rh{\mathcal{N}}_{0} = e^{i\theta}\cos{(\epsilon_{0}\rr{I_{C}})} \cos{(\epsilon_{0}\rr{I_{S}})}
\end{equation}

\subsection{Noisy beam splitter}
For the beam splitter we consider modes 0 and 1 as physical modes and mode 2 as virtual mode. The Hamiltonian operator we choose is that in Eq. \eqref{phase_H} with $\phi = 0$, ie $\rh{B}_{01}=\hat{a}_{0}\hat{a}_{1}^{\dagger} + \hat{a}_{0}^{\dagger}\hat{a}_{1}$. As Lindblad operators we choose $\rh{B}_{02} = \hat{a}_{0}\hat{a}_{2}^{\dagger} + \hat{a}_{0}^{\dagger}\hat{a}_{2}$ and $\rh{B}_{12} = \hat{a}_{1}\hat{a}_{2}^{\dagger} + \hat{a}_{1}^{\dagger}\hat{a}_{2}$. The corresponding noisy beam splitter reads
\begin{equation}\label{noisy_beam_splitter}
    \rh{N}_{012}=\rh{U}_{01}e^{i\epsilon_{0}\rh{S}_{02} + i\epsilon_{1}\rh{S}_{12}},
\end{equation}
where $\rh{U}_{01} = e^{i\frac{\theta}{2}\rh{B}_{01}}$ and $\rh{S}_{j2} = \int_{t_{0}}^{t}\rr{d\rr{W}_{j}(s)}\rh{B}_{j2}(s)$ with $j = 0,1$. In Eq. \eqref{noisy_beam_splitter} the parameters $\epsilon_{j}$ are defined as $\epsilon_{j} = \sqrt{-\log{(1-2\rr{p}_{j})}/2}$ analogously to the noisy phase shifter case. By performing a similar calculation to Eq.\eqref{B_phase_shifter_time} where now one uses the beam splitter relations in Eqs. \eqref{beam_relation_1},\eqref{beam_relation_2},\eqref{beam_relation_3} and \eqref{beam_relation_4} with the adjoint unitary evolution, the expressions for $\rh{B}_{j2}(s)$ are
\begin{align}
&\rh{B}_{02}(s) = \cos{\biggl(\frac{\theta(s,t_{0})}{2}\biggr)}\rh{B}_{02} - \sin\bigg({\frac{\theta(s,t_{0})}{2}}\bigg)\rh{C}_{12}, \\
& \rh{B}_{12}(s) = \cos{\biggl(\frac{\theta(s,t_{0})}{2}\biggr)}\rh{B}_{12} - \sin\bigg({\frac{\theta(s,t_{0})}{2}}\bigg)\rh{C}_{02},
\end{align}
where $\rh{C}_{j2} = i(\hat{a}_{j}^{\dagger}\hat{a}_{2} - \hat{a}_{j}\hat{a}_{2}^{\dagger})$.
The stochastic operators are then: 
\begin{align}
&\rh{S}_{02} = \rr{I}_{C0}\rh{B}_{02} - \rr{I}_{S0}\rh{C}_{12},\\
&\rh{S}_{12} = \rr{I}_{C1}\rh{B}_{12} - \rr{I}_{S1}\rh{C}_{02},
\end{align}
where $\rr{I}_{Cj} = \int_{t_{0}}^{t}\rr{d\rr{W}_{j}(s)}\cos{\theta(s,t_{0})}$ and $\rr{I}_{Sj} = \int_{t_{0}}^{t}\rr{d\rr{W}_{j}(s)}\sin{\theta(s,t_{0})}$. The latter Gaussian processes have the same variances and covariances computed before for the case of phase shifter. Finally we can express Eq.\eqref{noisy_beam_splitter} as 
\begin{equation}\label{noisy_beam_splitter_2}
    \rh{N}_{012}=\rh{U}_{01}e^{i\epsilon_{0}\rr{I}_{C0}\rh{B}_{02}}e^{-i\epsilon_{0}\rr{I}_{S0}\rh{C}_{12}}e^{i\epsilon_{1}\rr{I}_{C1}\rh{B}_{12}}e^{-i\epsilon_{1}\rr{I}_{S1}\rh{C}_{02}}\, ,
\end{equation}
and the corresponding action on creation operators is given by
\begin{equation}
\label{N_beam}
\medmath{
\begin{aligned}
&\rh{\mathcal{N}}_{012} = \begin{pmatrix} \cos{\theta/2} & i\sin{\theta/2}& 0\\
i\sin{\theta/2}& \cos{\theta/2}& 0\\
0 & 0 & 1
\end{pmatrix}\cdot
\begin{pmatrix} \cos{(\epsilon_{0}\rr{I_{C0}})} & 0 & i\sin{(\epsilon_{0}\rr{I_{C0}})}\\
0 & 1 & 0\\
i\sin{(\epsilon_{0}\rr{I_{C0}})} & 0 & \cos{(\epsilon_{0}\rr{I_{C0}})}
\end{pmatrix}\cdot \\
&\begin{pmatrix} 
1 & 0 & 0 \\
0 & \cos{(\epsilon_{0}\rr{I_{S0}})} & -\sin{(\epsilon_{0}\rr{I_{S0}})}\\
0 & \sin{(\epsilon_{0}\rr{I_{S0}})}&\cos{(\epsilon_{0}\rr{I_{S0}})}
\end{pmatrix}\cdot
\begin{pmatrix}
1 & 0 & 0 \\
0 & \cos{(\epsilon_{1}\rr{I_{C1}})} & i\sin{(\epsilon_{1}\rr{I_{C1}})}\\
0 & i\sin{(\epsilon_{1}\rr{I_{C1}})}&\cos{(\epsilon_{1}\rr{I_{C1}})}
\end{pmatrix}\cdot \\
&\begin{pmatrix} 
\cos{(\epsilon_{1}\rr{I_{S1}})} & 0 & -\sin{(\epsilon_{1}\rr{I_{S1}})}\\
0 & 1 & 0\\
\sin{(\epsilon_{1}\rr{I_{S1}})} & 0 & \cos{(\epsilon_{1}\rr{I_{S1}})}
\end{pmatrix}\, .
\end{aligned}
}
\end{equation}
By tracing out the virtual mode we get the expression for the noisy beam splitter
\begin{equation}
\label{N_beam_trace}
\begin{aligned}
&\rh{\mathcal{N}}_{01} = \begin{pmatrix} \cos{\theta/2} & i\sin{\theta/2}\\
i\sin{\theta/2}& \cos{\theta/2}
\end{pmatrix}\cdot
\begin{pmatrix} \cos{(\epsilon_{0}\rr{I_{C0}})} & 0 \\
0 & 1
\end{pmatrix}\cdot \\
&\begin{pmatrix} 
1 & 0 \\
0 & \cos{(\epsilon_{0}\rr{I_{S0}})}
\end{pmatrix}\cdot
\begin{pmatrix}
1 & 0 \\
0 & \cos{(\epsilon_{1}\rr{I_{C1}})}
\end{pmatrix}\cdot 
\begin{pmatrix} 
\cos{(\epsilon_{1}\rr{I_{S1}})} & 0\\
0 & 1
\end{pmatrix}\, .
\end{aligned}
\end{equation}

\section{Imperfect single-photon sources, lossy optical guides and detection}\label{other_noises}
As explained in section \ref{noise_dual_rail}, photon generation,  propagation of photons inside optical guides and detection are also affected by errors.
We now describe how to account for such errors with our formalism.

\subsection{Imperfect single-photon sources}
As we explained in section \ref{noise_dual_rail} a quantum dot undergoing decoherence processes generates non perfect indistinguishable photons. In the case of dual rail encoding, the only relevant property is the photon path, i.e. through which optical guide the photon is traveling. Then, in this setup, the only way to understand whether two photons are distinguishable, if we consider only non-coherent errors, is when the associated qubit state is not completely pure. This can be described by considering depolarizing errors arising at the sources and inherited by the photons.

To model imperfect single-photon sources we apply at the beginning of the optical circuit a layer of fictitious optical elements to each mode that simulate the depolarizing errors. Such optical elements can be built by applying again the noisy gates formalism. Here we focus on single qubit depolarizing channel, but one could straightforwardly extend the approach to multi-qubit depolarizing channel.

Depolarizing for a single spin state can be described by the Lindblad operators $\rh{X}$,$\rh{Y}$ and $\rh{Z}$. We can map such operators to operators defined in terms of creation and annihilation operators on the physical modes, thus acting on photons of the circuit. We notice for example that by solving the Schr\"odinger equation with the beam splitter Hamiltonian in Eq.\eqref{beam_H} with $\phi = 0$ and $\theta = \pi$, one gets a unitary on the creation operators that acts as an $\rh{X}$ gate on a qubit in the dual rail encoding (see Eq.\eqref{U_beam}). Similarly a $\rh{Y}$ is obtained with $\phi = \pi/2$ and $\theta = \pi$. The operator $\rh{Z}$ is obtained by solving the Schr\"odinger equation with the phase shifter Hamiltonian in Eq.\eqref{phase_H} with $\theta = \pi$ applied to mode 1.
For these reasons, to describe depolarizing errors, we can use Eq.\eqref{SDE} without Hamiltonian operator and with the Lindblad operators $\rh{B}_{01}$, $\rh{C}_{01}$ and $\rh{P}_1$.
The resulting noisy gate is: 
\begin{equation}\label{noisy_source}
    \rh{N}^{(\text{dep})}_{01}=e^{i\epsilon_{d}\rh{S}^{(x)}_{01}}e^{i\epsilon_{d}\rh{S}^{(y)}_{01}}e^{i\epsilon_{d} \rh{S}^{(z)}_{01}},
\end{equation}
where $\rh{S}^{(x)}_{01} = \rh{B}_{01} \rr{W}_x(t,t_0)$, $\rh{S}^{(y)}_{01} = \rh{C}_{01} \rr{W}_y(t,t_0)$ and $\rh{S}^{(z)}_{01} = \rh{P}_{1} \rr{W}_z(t,t_0)$. The corresponding unitary on the creation and annihilation operators becomes:
\begin{equation}
\label{N_dep}
\begin{aligned}
&\rh{\mathcal{N}}^{(\text{dep})}_{01} =
\begin{pmatrix} \cos{(\epsilon_{d}\rr{W}_x)} & i\sin{(\epsilon_{d}\rr{W}_x)}\\
i\sin{(\epsilon_{d}\rr{W}_x)} & \cos{(\epsilon_{d}\rr{W}_x)}
\end{pmatrix}\cdot \\
&\begin{pmatrix} 
\cos{(\epsilon_{d}\rr{W}_y)} & -\sin{(\epsilon_{d}\rr{W}_{y})}\\
\sin{(\epsilon_{d}\rr{W}_{y})}&\cos{(\epsilon_{d}\rr{W}_{y})}
\end{pmatrix}\cdot
\begin{pmatrix}
1 & 0 \\
0 & e^{i\epsilon_{d}\rr{W_{z}}} 
\end{pmatrix}.
\end{aligned}
\end{equation}
Here the parameter $\epsilon_{d}$ has the same definition of the loss parameter but now it is related to depolarizing probability $\rr{p}_{d}$. Moreover we stress that, despite the similarity with the noisy gate of lossy optical elements, the effects of Eqs. \eqref{noisy_source}, \eqref{N_dep} are different because modes 0 and 1 are both physical modes, there are not virtual modes.

\subsection{Lossy optical guides and detection}\label{lossy_optical_guides_and_detection}
In Sec. \ref{noisy_optical_elements} we have dealt with optical elements affected by photon losses, however photons inside optical guides and detectors are also subject to this kind of noise. Then, as in the case of noisy phase shifter we can consider two photon modes 0, 1, respectively physical and virtual mode, with Lindblad operator $\rh{B}_{01}$ but this time we use Eq.~\eqref{SDE} without Hamiltonian operator. A straightforward calculation leads to:
\begin{equation}\label{free_loss}
    \rh{N}^{(\text{loss})}_{01}=e^{i\epsilon\rh{S}_{01}},
\end{equation}
where $\rh{S}_{01} = \rh{B}_{01} \rr{W}(t,t_0)$ and $\epsilon = \sqrt{-\log{(1-2\rr{p})}/2}$ with $\rr{p}$ the loss probability inside the optical guide or the detector. Consequentially we have:
\begin{equation}
\label{N_loss}
\rh{\mathcal{N}}^{(\text{loss})}_{01} =
\begin{pmatrix} \cos{(\epsilon\rr{W})} & i\sin{(\epsilon\rr{W})}\\
i\sin{(\epsilon\rr{W})} & \cos{(\epsilon\rr{W})}
\end{pmatrix}.
\end{equation}
By tracing out the virtual mode we obtain:
\begin{equation}
\label{N_loss_trace}
\rh{\mathcal{N}}^{(\text{loss})}_{0} = \cos{(\epsilon\rr{W})}.
\end{equation}

\section{Comparison with other approaches}\label{comparisons}
In the literature, there are two main approaches for simulating photon losses: fixed-loss model \cite{aaronson2016bosonsampling,oszmaniec2016random,oszmaniec2018classical} and beam splitter loss model \cite{oszmaniec2018classical,brod2020classical,heurtel2023perceval,killoran2019strawberry}. The former one was developed when the number of lost particles can be effectively controlled; indeed it is based on the assumption that initially there are $n$ photons and exactly $n - l$ of them are lost. The loss is simulated by tracing out $n-l$ of the $n$ photons $\hat{\rho}' = \Tr_{n-l}(\hat{\rho})$ and by using the fact that $\Tr_{n-l}(\rh{U}^{\otimes n}\hat{\rho}\rh{U}^{\dagger \otimes n}) = \rh{U}^{\otimes n}\Tr_{n-l}(\hat{\rho})\rh{U}^{\dagger \otimes n}$. Thus, it does not matter whether losses occurred before or after the implemented linear transformation. However, a simulation method that describes a fixed number of lost photons is limited since in real optical circuits this number is a random variable.

In the second approach, the beam splitter loss model is based on the idea of adding a virtual mode to the physical mode and applying a beam splitter (see Eq. \eqref{U_beam} with $\phi = 0$) with a reflectance equal to the loss probability: $\rr{R} = \rr{p}$. Since the relation between reflectance and the angle $\theta$ of a beam splitter is $\cos(\theta/2) = \sqrt{1-\rr{R}}$ and $\sin(\theta/2) = \sqrt{\rr{R}}$, then one gets
\begin{equation}
\label{perceval_loss}
\begin{pmatrix} 
\sqrt{1-\rr{p}} & i\sqrt{\rr{p}}\\
i\sqrt{\rr{p}}& \sqrt{1-\rr{p}}
\end{pmatrix},
\end{equation}
where the probability of losing a photon is exactly $| _{\text{\tiny out}}\!\bra{01}\ket{10}_{\text{\tiny in}}|^{2} = \rr{p}$. This approach is very close to what we derived for lossy optical guides and detection in Sec. \ref{lossy_optical_guides_and_detection}, because Eq. \eqref{N_loss} is the stochastic version of Eq.~\eqref{perceval_loss}. Indeed, in our case the probability of losing a photon is $| _{\text{\tiny out}}\!\bra{01}\ket{10}_{\text{\tiny in}}|^{2} = \sin^{2}(\epsilon\rr{W})$ and by taking the average over the Wiener process one has
$\mathbb{E}[\sin^{2}(\epsilon\rr{W})] = (1-e^{-2\epsilon^{2}})/2 = \rr{p}$. 

This analogy is limited only to lossy optical guides and detectors. Indeed, the stochastic processes appearing in the expressions for noisy optical elements in Eq. \eqref{N_phase} and \eqref{N_beam} are not independent from the parameters of the corresponding optical elements. Therefore our approach is not reducible to the beam splitter loss model: as a result of including the noise into the gates, each noisy optical elements has its own lossy behaviour (even if the error probability is the same for all elements). Experimentally the dependence of photon losses on the parameters of optical elements is small and the usual assumption to perform simulations is that photon loss is independent from the angles of beam splitters and phase shifters \cite{mezher2022assessing}. However, we already proved in \cite{dibartolomeo2023novel}, for the case of superconducting devices, that relaxing this assumption, namely that of separating the noise and unitary evolution, provides a more accurate noise simulation tool. 

Regarding the treatment of imperfect photon sources, different approaches exist in the literature. In \cite{osca2023implementation}, the authors considered a parametrized quantum dot able to simulate the creation of pairs of entangled photon in a mixed state. Another example is the modeling used in \cite{Perceval} where an imperfect quantum-dot based single-photon source is modeled by a statistical mixture of Fock states. In this work we deal with single-photon sources thus, following the latter approach, we implemented the stochastic unraveling of a single qubit depolarizing Kraus map as shown in Eq.~\eqref{N_dep}, that equivalently outputs a mixed Fock state of the qubits. In general one could use $n$ qubits depolarizing channels to take into account multi photon distinguishability \cite{pont2022quantifying}.

\section{Testing the protocol}\label{tests}
In the following section we test the performances of our noisy gates
method. First, in subsection \ref{X_bell} we simulate the effects of imperfect photon sources and photon loss on the X gate and on the preparation of the Bell state in the gate based quantum computing (GBQC) framework. Then, in subsection \ref{X_CX} we perform the same simulation in the measurement based quantum computing (MBQC) framework \cite{raussendorf2001one,briegel2009measurement} reproducing the X gate. Finally in subsection \ref{vqa} we build a noisy variational quantum algorithm (VQA) \cite{cerezo2021variational,peruzzo2014variational} to solve the max 2-cut problem \cite{haastad2001some,berman1999some,proietti2022native} and we study its performances in different noise scenarios. 
\begin{figure*}[htp]
\begin{minipage}{0.45\textwidth}
\includegraphics[width=\textwidth]{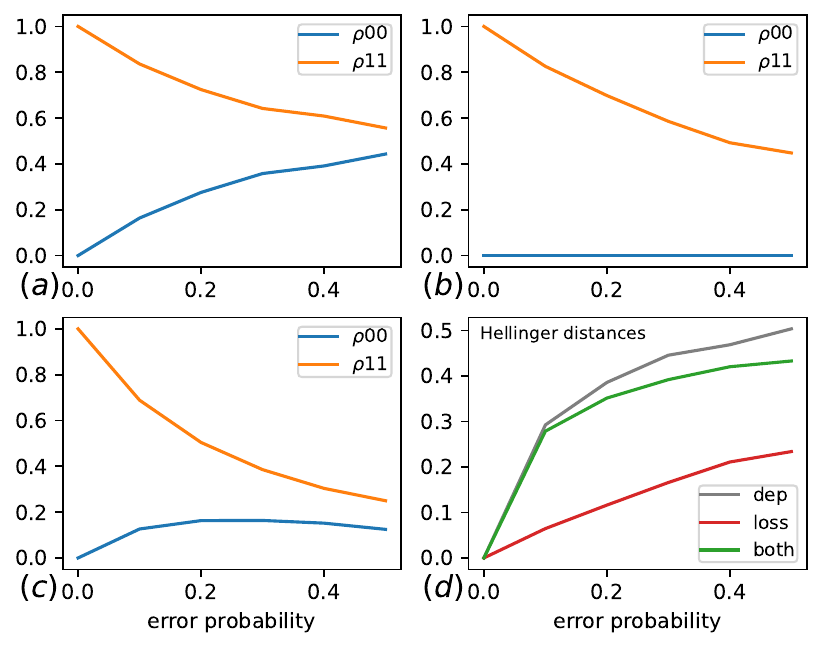}
\end{minipage}
\begin{minipage}{0.30\textwidth}
\includegraphics[width=0.8\textwidth]{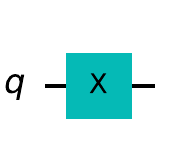}\\
\includegraphics[width=\textwidth]{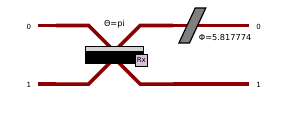}   
\end{minipage}
\caption{X gate in the gate based framework. Panels (a), (b), (c) show the values of the entries $\rho_{00}$ (blue) and $\rho_{11}$ (orange)  of the density matrix for the X gate, as the error probabilities increase. Panel (a) takes into account only imperfect photon sources (modeled as depolarization), panel (b) photon loss errors and panel (c) the combination of the two. Panel (d) shows the Hellinger distances between the outcomes of the simulations in (a) (grey), (b) (green), (c) (red) and the ideal result of the X gate. 
On the right, the circuit for the X gate in GBQC and the corresponding optical circuit.}
\label{plots_X_GBQC}
\end{figure*}

All the optical circuits we implement are written in the standard triangular Reck decomposition \cite{reck1994experimental}, except for the VQA ansatz. As computational backend for the simulations we used the SLOS backend \cite{heurtel2023strong} in Perceval \cite{Perceval} and measurements in MBQC framework are handled with the MBQC module of Paddle Quantum \cite{Paddlequantum}. Moreover, the transpilation required to obtain the optical circuits in subsections \ref{X_bell} and \ref{X_CX} are performed by using the QiskitConverter functionality of Perceval.

\begin{figure}[htp]
\includegraphics[width=0.50\textwidth]{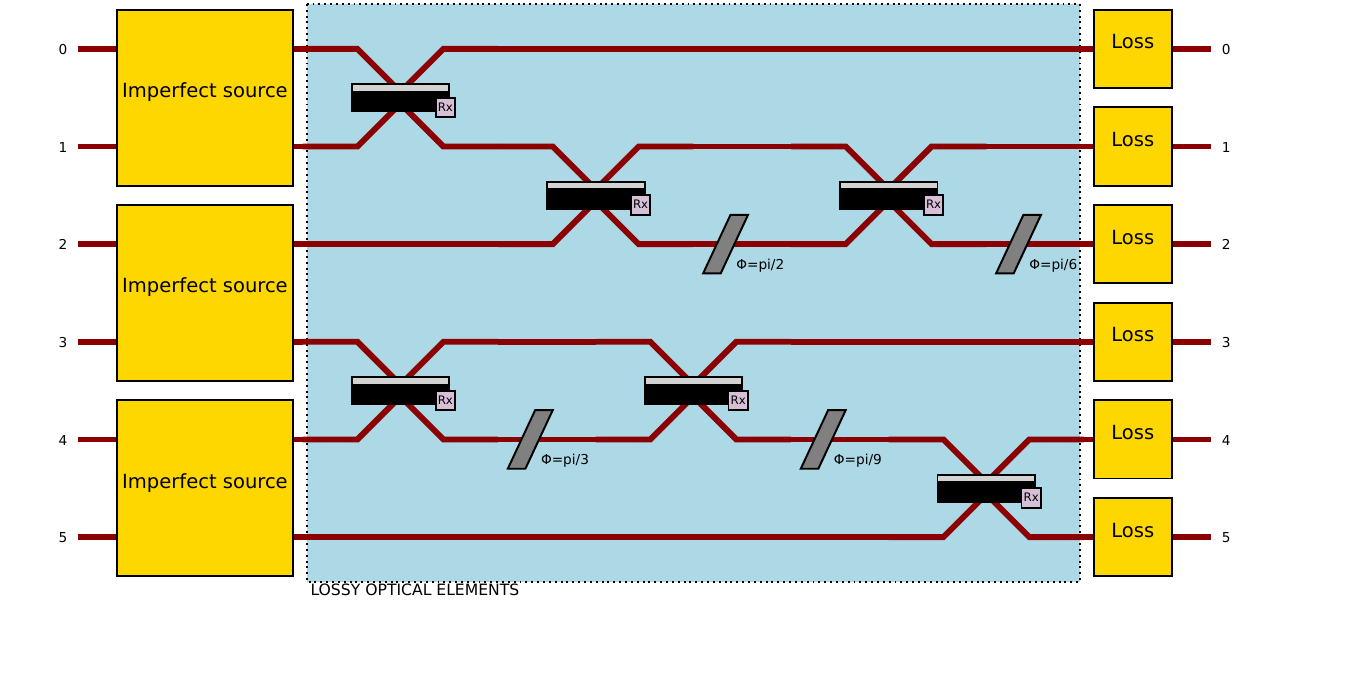}
\caption{Schematic depiction of how we simulate a generic noisy optical circuit. On the left in yellow, a layer of fictitious optical elements models depolarization for each couple of modes. In the blue area, each optical element of the original circuit is replaced with the corresponding noisy one and for losses at detection we add a layer of lossy channel for each mode, shown here in yellow on the right.}
\label{noisy_circuit_example}
\end{figure}

To simulate the effect of imperfect photon sources, we add to the optical circuit a layer of fictitious optical elements modeling depolarization, see Eq.~\eqref{N_dep}, for each couple of modes. To simulate photon losses, each optical element of the original circuit is replaced with the corresponding noisy one, see Eqs.~\eqref{N_phase_trace} and~\eqref{N_beam_trace}, and for losses at detection we add a layer of lossy channel for each mode as in Eq.~\eqref{N_loss_trace}. The corresponding noisy optical circuit is schematically depicted in Fig. \ref{noisy_circuit_example}.

\subsection{X gate and Bell state in the gate based framework}\label{X_bell}
We simulate the X gate and Bell state preparation with our approach, eventually obtaining the output density matrix $\rho^{\text{ng}}$. If $\sigma$ is the noiseless output density matrix,  we compare these two states by computing the Hellinger distance $\cl{H}^{\text{ng}} = \cl{H}(\rho^{\text{ng}},\sigma)$, where the Hellinger distance is defined by
\begin{equation}
    \cl{H}(\rho,\sigma)=\frac{1}{\sqrt{2}}\sqrt{\sum_{k=1}^{N}\big(\sqrt{\rho_{kk}}-\sqrt{\sigma_{kk}}\big)^2}\, ,
\end{equation} with $\rho_{kk}$ ($\sigma_{kk}$) the diagonal elements of $\rho$ ($\sigma$). The Hellinger distance is a measure of the distance between the readout probability distributions, and is related to the fidelity as $\cl{F} = (1-\cl{H}^2)^2$ \cite{nielsen2000quantum}. We chose to work with the Hellinger distance since the fidelity is not a metric, and we are interested in the distance between readout probability distributions. We repeat these simulations for diverse type of noise (depolarizing and photon loss) and different error probabilities.

\begin{figure*}[htp]
\centering
\begin{minipage}{0.45\textwidth}
\includegraphics[width=\textwidth]{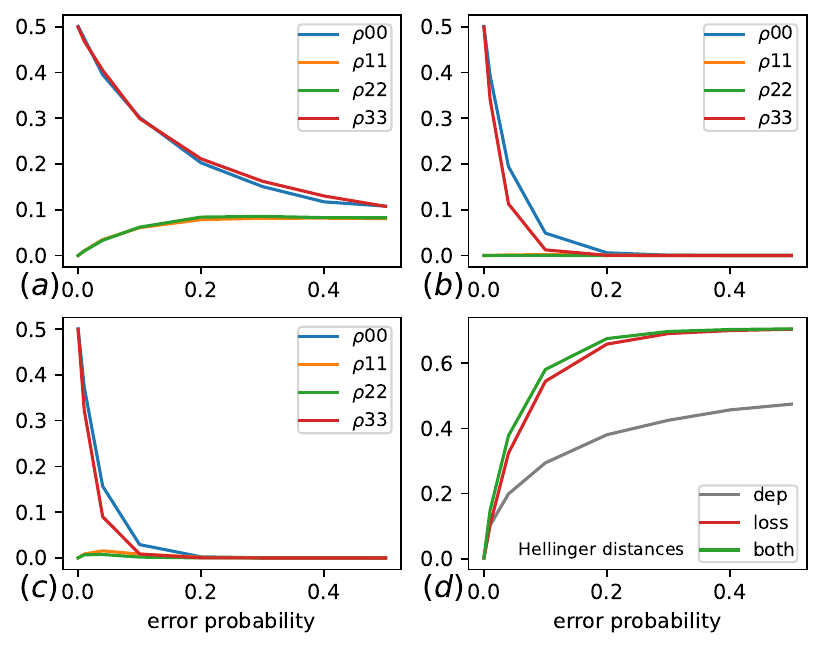}
\end{minipage}
\begin{minipage}{0.30\textwidth}
\includegraphics[width=\textwidth]{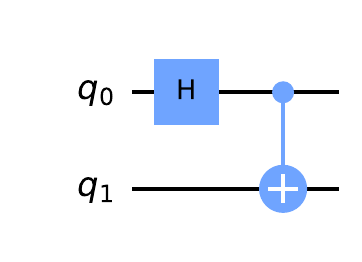}
\end{minipage}
\begin{minipage}{0.95\textwidth}
\includegraphics[width=\textwidth]{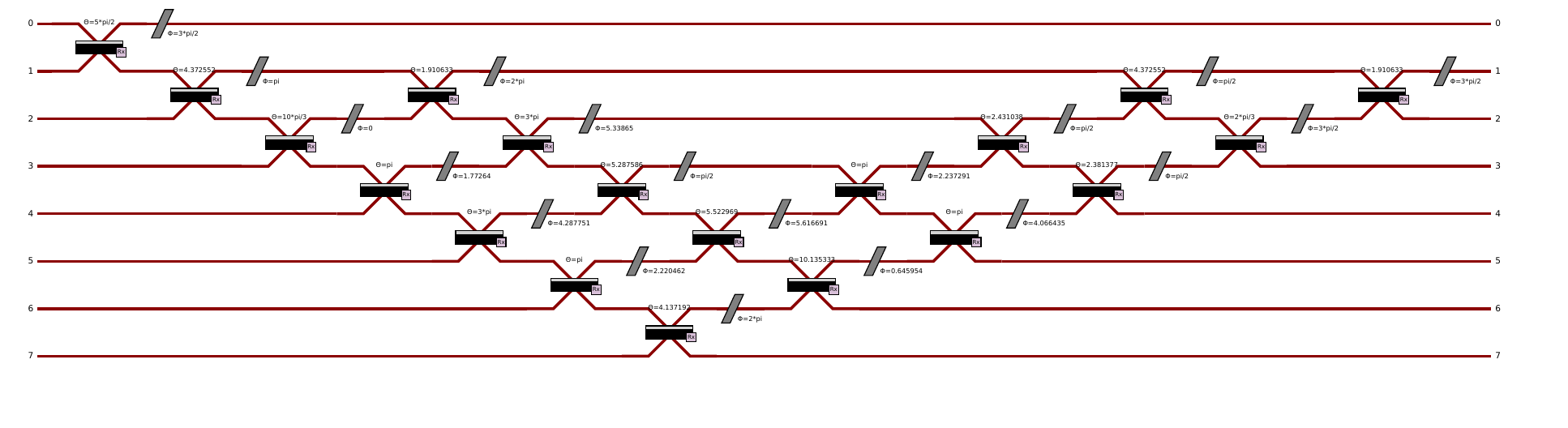}
\end{minipage}
\caption{Bell state in the gate based framework. Panels (a), (b), (c) show the values of the entries of $\rho_{00}$ (blue) and $\rho_{11}$ (orange), $\rho_{22}$ (green), $\rho_{33}$ (red)  of the density matrix for the Bell state. The order of the different noise sources considered is the same as for the X gate Fig.\ref{plots_X_GBQC}. Panel (d) shows the Hellinger distances between the outcomes of the simulations and the ideal result of the Bell state, colors have the same meaning as for panel (d) of Fig.\ref{plots_X_GBQC}. On the right the circuit for the Bell state in GBQC is shown, and on the bottom the corresponding optical circuit.}
\label{plots_bell_GBQC}
\end{figure*}

The results of the simulations are shown in Figs. \ref{plots_X_GBQC} and \ref{plots_bell_GBQC}. All results shown are normalized with respect to the postselection/heralding probability. In Fig. \ref{plots_X_GBQC}, panel (a) shows the effect of depolarization that is to bring the system to the completely mixed state, with $\rho_{00} = \rho_{11} = 1/2$. The effect of photon losses, shown in panel (b), is to decrease $\rho_{11}$, as expected: in this case the state is not normalized as the action of losses is non unitary, see also Eqs.~\eqref{N_phase_trace}, \eqref{N_beam_trace} and \eqref{N_loss_trace}. The combined action of depolarization and photon losses is shown in panel (c): the effects of depolarization are lowered by the presence of photon losses. Panel (d) displays the Hellinger distances of the three simulations with respect to the ideal result. 
For the circuit preparing the Bell state in Fig. \ref{plots_bell_GBQC}, the effect of depolarization, shown in panel (a), is again to bring the system to a mixed state. In general, for two qubits the equal probability of the maximally mixed state is $1/4$, however in this case depolarization is applied not only on system qubits but also on ancillary qubits, necessary to implement the CX gate; this leads to a mixed state different from the maximally mixed state for two qubits. 

The effect of photon losses in panel (b) is way larger than for the X gate; the reason is that  the optical circuit to realize the Bell state has a significantly higher depth and a larger number of modes as can be seen by comparing the optical circuits in Figs. \ref{plots_X_GBQC} and \ref{plots_bell_GBQC}. Panel (c) shows that photon losses are dominant in this case, as the plot is very close to panel (b). Panel (d) displays the Hellinger distances of the three simulations with respect to the ideal result.
\begin{figure*}[htp]
\centering
\begin{minipage}{0.45\textwidth}
\includegraphics[width=\textwidth]{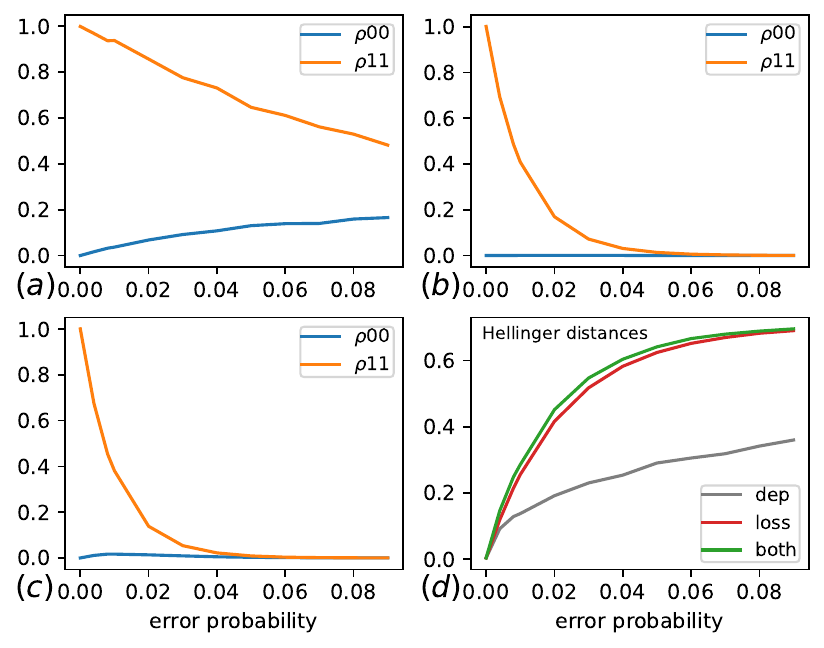}
\end{minipage}
\begin{minipage}{0.30\textwidth}
\includegraphics[width=\textwidth]{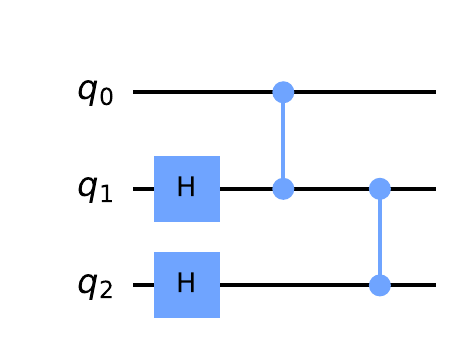}
\end{minipage}
\begin{minipage}{0.95\textwidth}
\includegraphics[width=\textwidth]{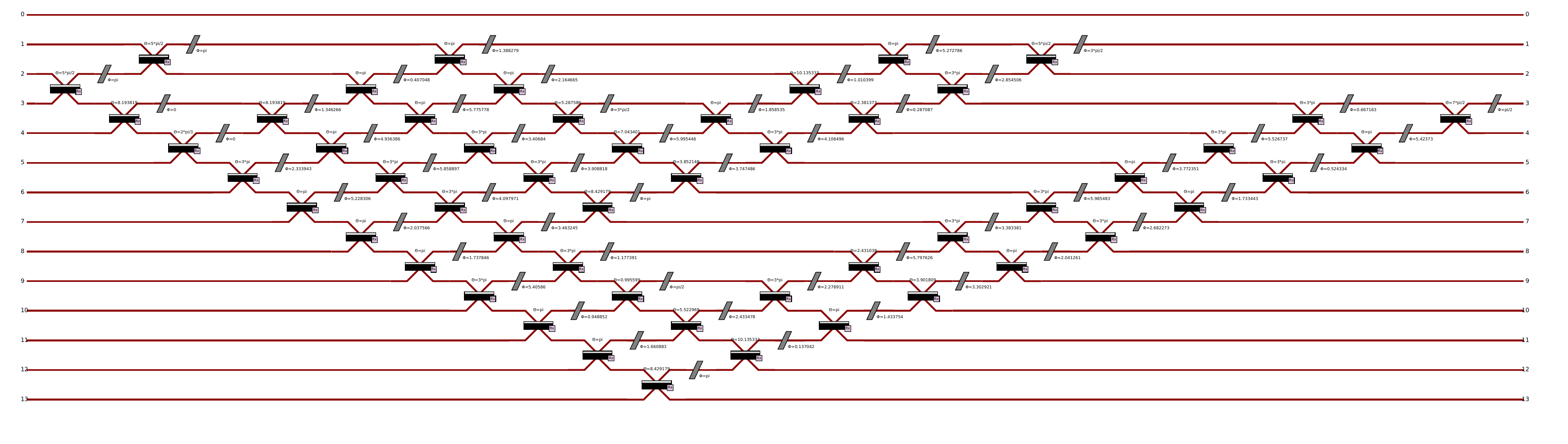}
\end{minipage}
\caption{X gate in the measurement based framework. Panels (a), (b), (c) show the values of the entries $\rho_{00}$ (blue) and $\rho_{11}$ (orange)  of the density matrix for the X gate in the MBQC framework as the error probabilities increase. Panel (a) takes into account only imperfect photon sources (modeled as depolarization), panel (b) photon loss errors and panel (c) the combination of the two. Panel (d) shows the Hellinger distances between the outcomes of the simulations in (a) (grey), (b) (green), (c) (red) and the ideal result of the X gate. On the right, the circuit required to prepare the cluster state for the X gate in MBQC is shown, and at the bottom the corresponding optical circuit is displayed. Measurements for the MBQC are not shown.}
\label{plots_X_CX_MBQC}
\end{figure*}

\subsection{X gate in the measurement based framework}\label{X_CX}
\begin{figure*}[htp]
    \begin{minipage}{0.330\textwidth}
        \includegraphics[width=\textwidth]{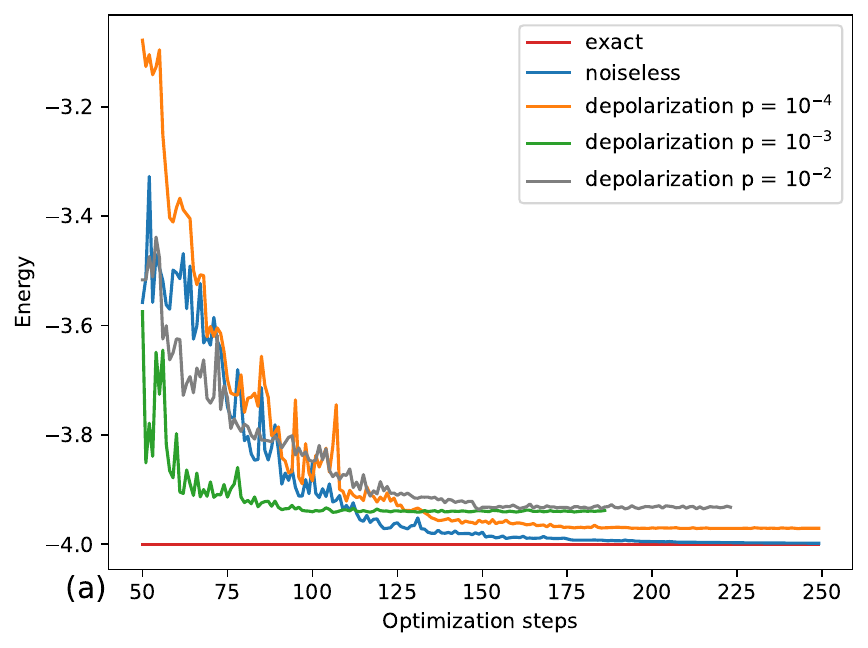}
    \end{minipage}%
    \hfill%
    \begin{minipage}{0.330\textwidth}
        \includegraphics[width=\textwidth]{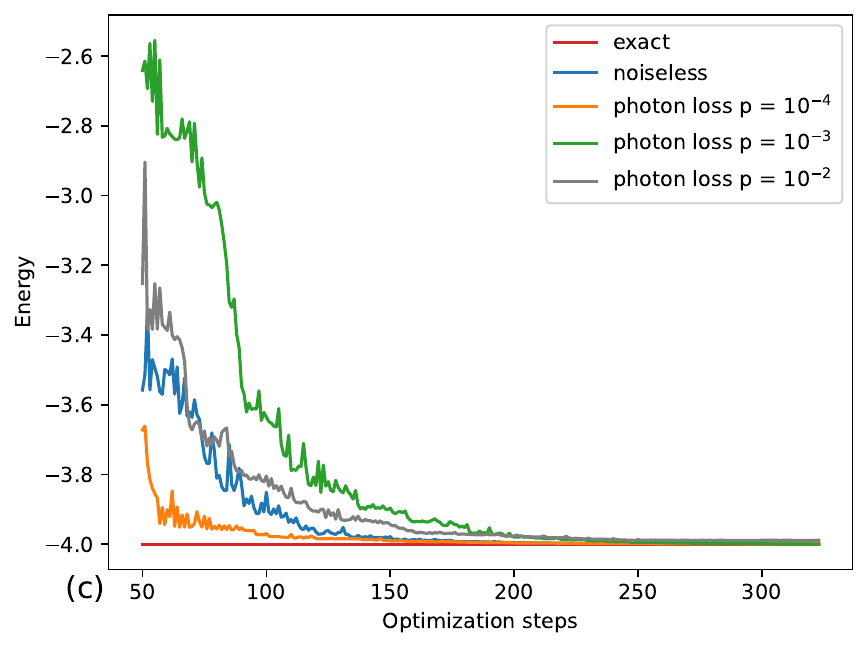}
    \end{minipage}
    \hfill%
    \begin{minipage}{0.330\textwidth}
        \includegraphics[width=\textwidth]{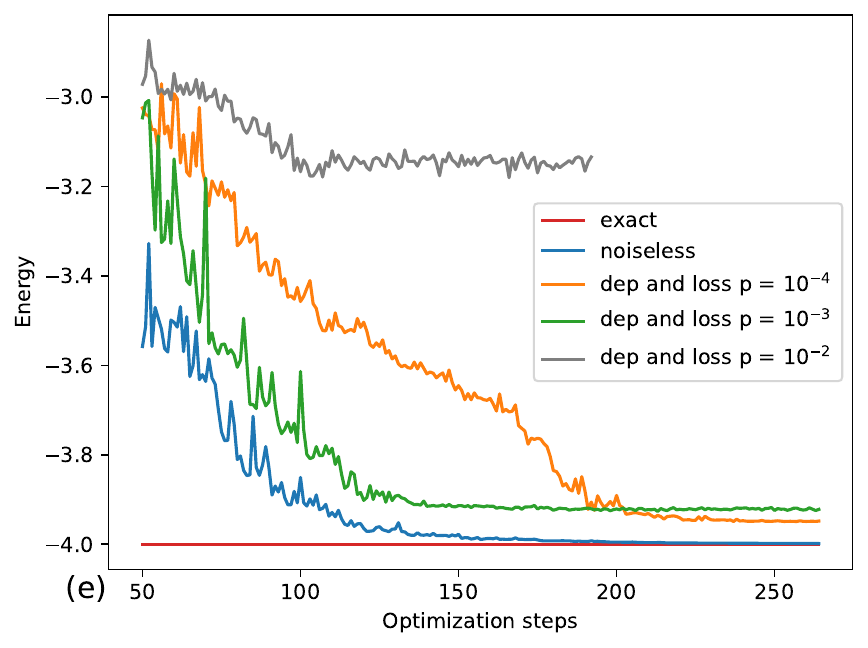}
    \end{minipage}
    \begin{minipage}{0.330\textwidth}
        \includegraphics[width=\textwidth]{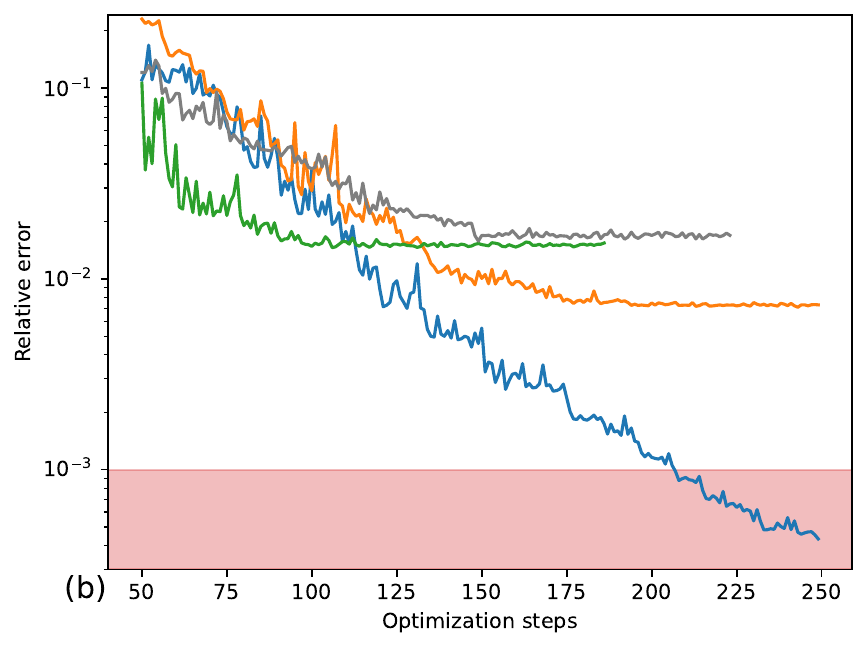}     
    \end{minipage}%
    \hfill%
    \begin{minipage}{0.330\textwidth}
        \includegraphics[width=\textwidth]{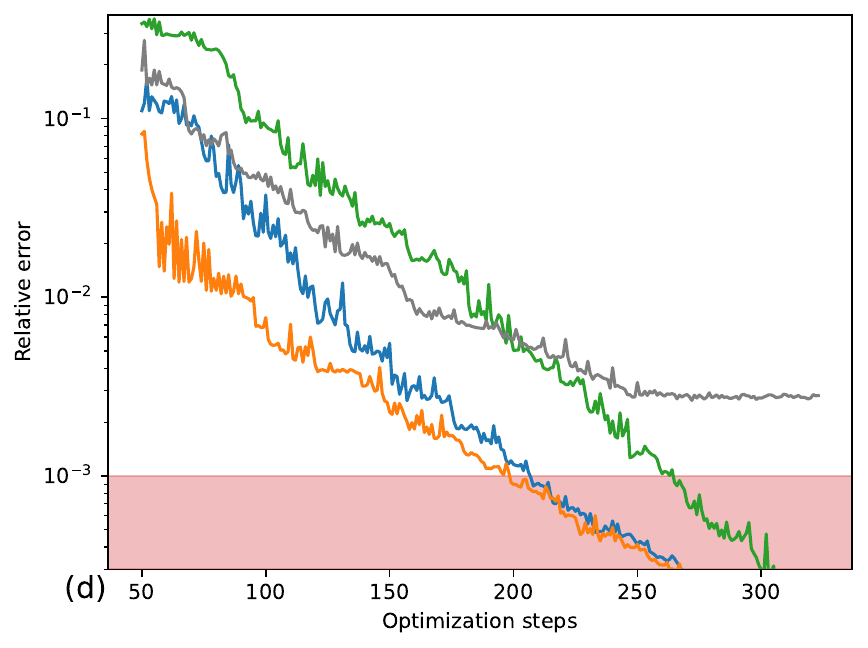}      
    \end{minipage}
    \hfill%
    \begin{minipage}{0.330\textwidth}
         \includegraphics[width=\textwidth]{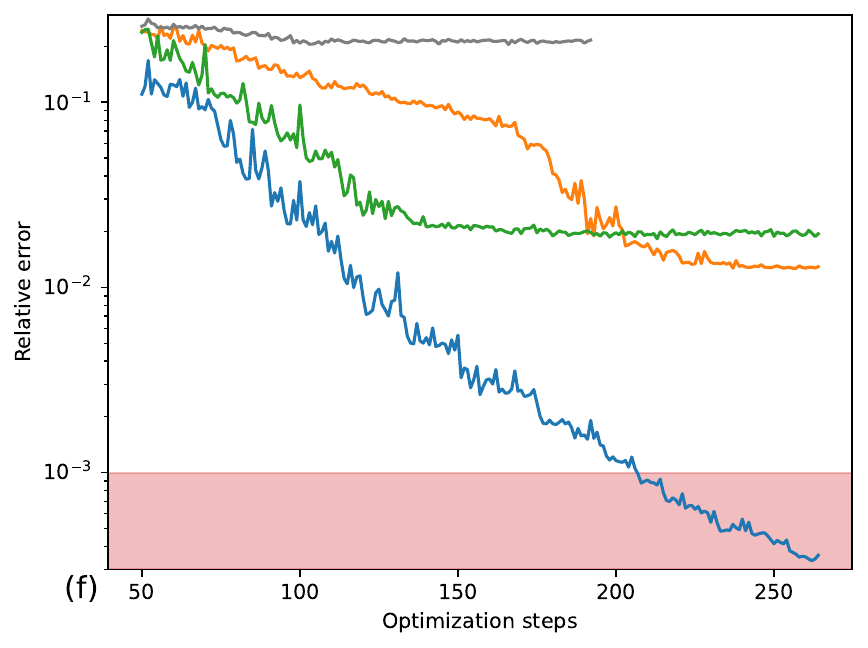}
    \end{minipage}
    \begin{minipage}{0.95\textwidth}
    \centering
        \includegraphics[width=\textwidth]{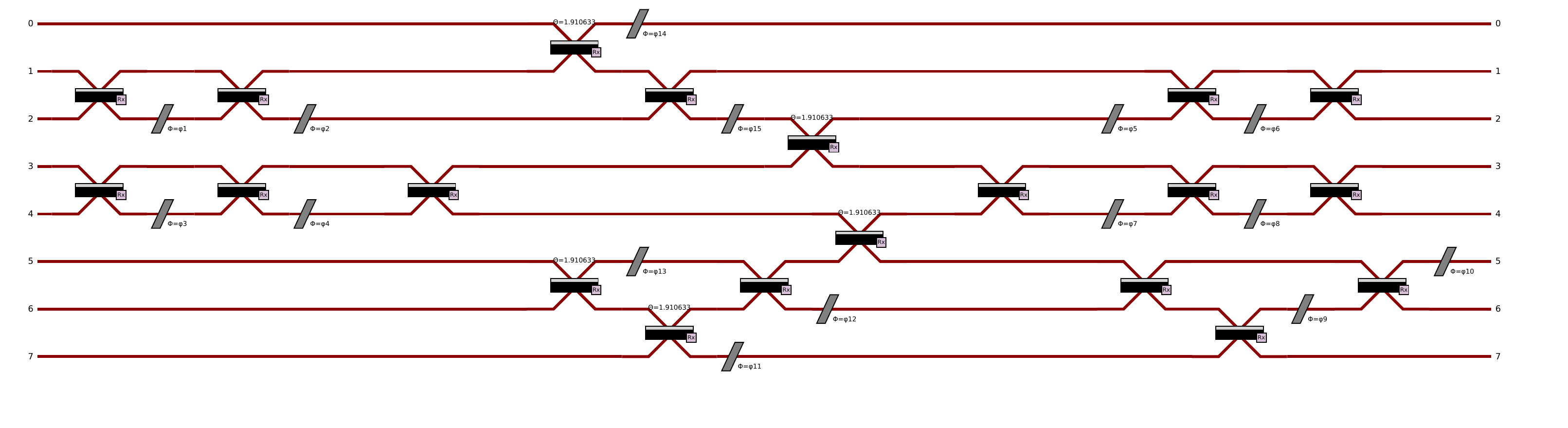}
    \end{minipage}
\caption{Panels (a), (c), and (e) show the optimization of the variational energy as a function of the number of optimization steps obtained with the simulations. Each curve is obtained by simulating the variational ansatz starting from random parameters. For visualization purposes, all panels start from the 50th optimization step. In each (a), (c) and (e) panels the red line is the exact energy $E_{0}= -4$, the blue curve is the optimization in the noiseless case (which is the same curve in all plots), the orange, green and grey curves are obtained with an error probability of $10^{-4}$, $10^{-3}$, $10^{-2}$ respectively. Figures (b), (d), and (f) show the relative errors $|(E_{0} - E)/E_{0}|$ as a function of the training steps. Colors of the curves have the same meaning of the upper panels. A $0.1\%$ relative error threshold is here highlighted with a red shaded region for better visualization. On the bottom, the optical circuit ansatz for the VQA solving the max 2-cut on a square graph are shown.}
\label{energies_simulations}
\end{figure*}
The framework of dual rail encoding of linear optics quantum computing is suitable to implement alternative scheme for quantum computation, in particular measurement based quantum computing (MBQC). Such a scheme involves the generation of particular entangled states called cluster states and appropriate single qubit measurement to drive the computation. These two stages involve photon generation, manipulation and detection thus, as for GBQC, MBQC is affected by photon losses and imperfect photon sourcing. To test the robustness of MBQC against such noises, we focus on the realization of X gate.
The simplest implementation of the X gate is schematically depicted as
\begin{equation}
\label{Xcluster}
\begin{quantikz}
\lstick{$\ket{0}$}&\qw&\ctrl{1}&\qw&\meter{}& \rstick{$X = s_1$}\\
\lstick{$\ket{0}$}&\gate{H}&\ctrl{0}&\ctrl{1}&\meter{} \rstick{$R_{XY}((-1)^{1+s_1}\pi) = s_2$}\\
\lstick{$\ket{0}$}&\gate{H}&\qw&\ctrl{0}&\qw \rstick{$X^{s_2}Z^{s_1}X\ket{0}$}
\end{quantikz}
\end{equation}
Qubit 1 is prepared in the state $\ket{0}$, while qubits 2,3 are both prepared into the $\ket{+}$ state. Qubit 1 is entangled with 2, and qubit 2 is entangled with 3 via CZ gates; The output state at this stage is called cluster state. Then the first qubit is measured in the X basis and the measurement outcome $s_1$ is recorded. Qubit $2$ is measured in the XY plane with an angle $\theta = (-1)^{1+s_1}\pi$ and the measurement outcome $s_2$ is recorded. After these two measurements, the  state of qubit 3 is $\rh{X}^{s_2}\rh{Z}^{s_1}\rh{X}\ket{0} = \rh{X}^{s_2}\rh{Z}^{s_1}\ket{1}$. Thus the initial $\ket{0}$ is flipped to $\ket{1}$ up to byproduct Pauli operators. We notice that a generic rotation around the X axis of angle $\alpha$ is realized by taking $\theta = (-1)^{1+s_1}\alpha$. The corresponding optical circuit that implements the cluster state for the X gate and the effect of noise are shown in Fig. \ref{plots_X_CX_MBQC}. 

As in the previous section, the effect of depolarization is to bring the system to
the completely mixed state (panel (a)) and the effect of photon losses (panel (b)) is to decrease $\rho_{11}$; panel (c) reports the effect of both depolarization and photon losses. As one can notice by comparing the order of magnitude of the error probabilities of Figs. \ref{plots_X_GBQC}, \ref{plots_bell_GBQC} and  Fig. \ref{plots_X_CX_MBQC}, the impact of noise in the MBQC framework is way stronger due to the larger number of modes and optical elements in the MBQC optical circuit. For example, to perform the GBQC X gate, two modes and one beam splitter are enough, while for the MBQC X gate cluster state in the Reck decomposition one needs 14 modes and 116 optical elements.

We notice that all the CZ gates are implemented by adding four additional modes to herald the success of the gate. This extra ancillary cost is not always required, and most efficient layouts can be found but for the only purpose of testing our method, we did not further optimize the photonic circuits.

\subsection{Variational quantum algorithm applied to the max 2-cut problem}\label{vqa}
We now focus on a VQA algorithm applied to the MAX $2$-CUT problem. The MAX $2$-CUT problem is a well known optimization problem that requires to find a cut dividing the vertices of a graph in two complementary subsets, such that the number of edges crossed by the cut is maximized. Given a graph $G = (V,E)$, with $V = \{ 1,\dots,N \}$ and $E =\{ \langle i,j \rangle \}$ the sets of vertices $i$ and edges $ \langle i,j \rangle$, the goal is to maximize the following objective function
\begin{equation}
    \rh{H}_C = \sum_{\langle i,j \rangle} \frac{1}{2}(1-\rh{Z}_i\rh{Z}_j).
\end{equation}

We focus on a graph with $V = \{1,2,3,4\}$ and $E = \{ \langle 1,2 \rangle \, \langle 2,3 \rangle \, \langle 3,4 \rangle \, \langle 4,1 \rangle \}$, thus a square graph with four nodes. The problem is equivalent to minimizing the cost Hamiltonian:
\begin{equation}\label{cost_ham_ansatz}
    \rh{H}_C = \sum_{\langle i,j \rangle}\rh{Z}_i\rh{Z}_j;
\end{equation}
for this cost Hamiltonian, the optimal, or exact, energy is $E_{0} = -4$, corresponding to the cut giving $V_1= \{1,3\}$ and $V_2= \{2,4\}$ as complementary subsets of $V$. This is the target energy of the optimization of the variational ansatz.

We choose a specific ansatz $\rh{U}_{ansatz}(\theta)$ for the optimization shown in the bottom of Fig.~\ref{energies_simulations}.
 The output of each simulations is a density matrix $\hat{\rho}$, as shown in Eq. \ref{mean_rho}. The cost function for the optimization is the energy calculated as $E = \Tr(\hat{\rho}(\theta)\rh{H}_C)$, where $\hat{\rho}(\theta) = \ket{\psi(\theta)}\bra{\psi(\theta)}$ and $\ket{\psi(\theta)} = \rh{U}_{ansatz}(\theta)\ket{0}^{\otimes 4}$. We choose a number of samples $\rr{N}_{samples} = 500$ in Eq. \ref{mean_rho} and initial parameters are chosen randomly. The classical optimizer method used to update the parameters is the gradient-free method COBYLA \cite{COBYLA}. 

We study the convergence of the optimization in the noiseless case, in the presence of depolarization with $\rr{p} = 10^{-4},10^{-3},10^{-2}$, in the presence of loss with $\rr{p} = 10^{-4},10^{-3},10^{-2}$ and under the combined action of depolarization and loss with $\rr{p} = 10^{-4},10^{-3},10^{-2}$. We restrict the analysis to such orders of magnitude of the error probability, because for $\rr{p}<10^{-4}$ the effects of noises on the ansatz circuit are negligible and for $\rr{p}>10^{-2}$  results might not be reliable since our approach is perturbative, as shown in section \ref{formalism}, as the error probability approaches $1$. 

The results are shown in Fig. \ref{energies_simulations}. The noiseless optimization converges to the optimum value, showing that the ansatz works. 
Panels (a) and (b) show the simulations in the presence of depolarization: as expected, the higher the error probability, the higher the deviation from the exact energy. When $p = 10^{-4}$, the variational algorithm is still able to output an estimate $E$ with a $\sim 1\%$ relative error. 
In case of photon losses, shown in panels (c) and (d), the ansatz seem to be more resilient: for $\rr{p} = 10^{-4}$ and $\rr{p} = 10^{-3}$ the relative error is below $0.1\%$ and for $\rr{p}= 10^{-2}$ the relative error is below $1\%$. The combined action of photon losses and depolarization, shown in panels (e) and (f), is to further decrease the accuracy of the optimization as all relative errors are above $1\%$.

We report the approximation ratios, defined as $R = E_{f}/E_{0}$ where $E_{f}$ is the final energy for different noises and different orders of magnitude of the error probability in Table~\ref{table_approximation_ratio}. The best approximation ratios are obtained with the simulations considering photon losses only. 

The choice of an appropriate ansatz is a critical factor in determining the VQA model's performance \cite{VARIATIONAL_QUANTUM_ALGORITHMS}. The motivation behind opting for a carefully balanced ansatz lies in avoiding excessive expressibility, which can potentially lead to challenges during the classical optimization process. An ansatz with overly complex representations may be able to effectively capture the features of the problem's Hamiltonian. However, this increased expressibility can come at the cost of introducing a large number of parameters and intricate interactions. As a consequence, the cost function gradients may vanish exponentially during classical optimization \cite{du2022efficient}. This phenomenon hinders the optimization process, making it difficult for the model to converge efficiently and find satisfactory solutions. In the circuit model of quantum computing, several ansatz designs have been developed to construct parameterized quantum circuits tailored to different tasks and applications \cite{VARIATIONAL_QUANTUM_ALGORITHMS}.

\begin{table}[t!]\label{table_approximation_ratio}
\vspace{5mm}
\centering
\begin{tabular}{|c|c|c|c|}
\hline
& $\rr{p} = 10^{-4}$ & $\rr{p} = 10^{-3}$ & $\rr{p} = 10^{-2}$\\
\hline 
Depolarization & 0.9927 & 0.9846 & 0.9831 \\
\hline
Photon Loss & 0.999998 & 0.999985 & 0.997180\\
\hline
Dep and Loss & 0.9875 & 0.9808 & 0.7835\\
\hline
\end{tabular}
\caption{Values of the approximation ratio for different types of noises and for increasing order of magnitude of the error probability.}
\end{table}

\section{Conclusions and Outlook}
We have extended the noisy gates approach \cite{dibartolomeo2023novel} to simulate noisy optical circuits, within the second quantization framework (see sections \ref{noisy_optical_elements} and \ref{other_noises}), which is the one commonly used in the dual rail encoding for linear optics quantum computing. The simulation method is applicable both to GBQC and MBQC, as shown in sections \ref{X_bell} and \ref{X_CX}. In section \ref{vqa} we have shown that the simulations are suitable to describe variational problems when noises are present.

When considering quantum computing on linear optics, the situation is not yet as mature as for other quantum platforms. Linear optical quantum computing, which utilizes photons for quantum information processing, faces unique challenges in designing efficient and expressive ansatz circuits. The absence of non-linear interactions in linear optics limits the complexity of quantum computations that can be performed. Consequently, creating ansatz designs that can effectively encode and manipulate quantum information in linear optics remains an ongoing research challenge. MBQC was proposed as a solution to these issues. 

For this reason, the next natural target of this work is to understand the practical applicability of a noisy native MBQC algorithm, such as the Quantum Approximate Optimization Algorithm (QAOA) \cite{farhi2014quantum} formulated in the MBQC framework for the max K-cut problem in \cite{proietti2022native}. 

Another important goal is to compare our simulations with the output of optical devices available in the cloud, for example the one recently provided by Quandela \cite{maring2023general}. These targets will be the subject of future research.

\begin{acknowledgments}
\noindent
This work was funded by the call ``Solvers wanted" by Leonardo S.p.A. All the authors thank A. Salavrakos, J. Senellart, S. Mansfield and N. Somaschi for useful discussions. A.B., G.D.B. and M.V. acknowledge the  support from University of Trieste and INFN. M.V. thanks L. Ruscio for useful discussions.
\end{acknowledgments}



\bibliography{biblio}
\end{document}